\documentclass[preprint,12pt]{elsarticle}
\usepackage{amsmath}
\usepackage{amssymb}
\usepackage[pdftex,colorlinks]{hyperref}
\usepackage{multirow}
\usepackage{tabularx}
\usepackage{booktabs}
\usepackage{subfig}
\usepackage{graphicx}
\usepackage{dcolumn}
\usepackage{bm}
\biboptions{compress}

\begin{document}

\begin{frontmatter}

\title{Full quantum theory of control-not gate in ion-trap quantum computation}

\author{Biyao Yang $^{a}$}
\author{Li Yang$^{b}$\corref{1}}
\cortext[1]{ yangli@iie.ac.cn}
\address{a.National Secrecy Science and Technology Evaluation Center, Beijing 100044, China\\
b.Institute of
Information Engineering, CAS, Beijing 100093, China}

\begin{abstract}
We investigate the exact effect on ion trap quantum computation after field quantization. First an exact expression of failure probability from field quantization after many CNOT operations in Cirac-Zoller scheme is given. It is proportional to operation number and the amplitude of $|1\rangle_x
|0\rangle_y$ or $|1\rangle_x |1\rangle_y$ in initial state, and inverse
proportional to mean number of photons and amplitude of
$|0\rangle_x |0\rangle_y$ or $|0\rangle_x |1\rangle_y$ in initial state. Then we calculate the failure probability when the limitation to mean number of photons in sideband transition is considered. When the initial state is
$|1\rangle_x |0\rangle_y$ or
 $|1\rangle_x |1\rangle_y$, after about $10^2$ times of
CNOT operations, failure probability is no less than $10^{-2}$, while $10^{-2}$ is the known maximum threshold in fault-tolerant quantum computation. Then when the initial state is
$|1\rangle_x |0\rangle_y$ or
 $|1\rangle_x |1\rangle_y$, the number
of CNOT gates on the same pair of physical qubits should be no
more than $10^2$ in one error-correction period, or else the
computation cannot be implemented reliably. This conclusion can help to determine the number of CNOT operations between coding and decoding in one error-correction period in fault-tolerant quantum computation.\end{abstract}




\end{frontmatter}


\section{Introduction}

Quantum computers (QC) can solve some problems considered intractable on classical computers \cite{Shor1994,Grover1996}. The difficulty of these problems on classical computers is the foundation of some widely used public-key cryptosystems \cite{RSA1978,ElGamal1985}. Many physical realization schemes of QC have been proposed. Among them, the ones which are driven by coherent field are the most promising. Examples of this type of scheme are cold ion-trap scheme (Cirac-Zoller scheme) \cite{Cirac&Zoller1995}, cavity quantum electrodynamics scheme, etc. In particular, in the cold ion-trap QC, it has good scalability and small decoherence rate, and much effort has been made toward this direction \cite{Monroe1995,Schmidt-Kaler2003,Duan2006, Leibrandt2009}.

In the QC driven by coherent field, the field used for controlling the computation is usually taken as classical. Actually, the field is a quantum system itself, then when it is quantized strictly in the physical system of QC, results different from that of classical treatment of the field may arise. There are generally two ways to take into account the quantum nature of the driving field. One is to add quantum fluctuation to classical treatment of laser field \cite{Julio2002}. There are many operations in QC, then whether this method is valid in the case of quantum computation is still to be considered. The other method to consider the quantum nature of driving field is to strictly quantize the laser field and work out the exact result \cite{Yang07, Yang12}. In \cite{Yang07}, they combined the limitation given by the
quantum nature of the field with the threshold theorem in fault-tolerant quantum computation, and gave the concept of permitted depth of logical operation, which limits the number of operations on any physical qubit in one
error-correction period; in \cite{Yang12}, for Rabi oscillation of a two-level system driven by a quantized pulse train, one application of which is QC, they develop a model to deal with the system, give an algorithm to solve with any given precision the complicated infinite sums resulting from the model above, and get some meaningful results about this basic physical interaction. Then for ion-trap QC, they calculate the failure probability after a $k\pi$ pulse train, and estimate the upper bound of controlled-NOT (CNOT) operation number in Cirac-Zoller scheme.

There are actually five steps in Cirac-Zoller CNOT scheme, but in \cite{Yang12}, they only calculate failure probability of one step, and take it as the lower bound of the failure probability after one CNOT operation. Then the failure probability after repeated CNOT operations they calculate is smaller compared to the exact one, resulting a larger upper bound for operation number. In this paper, we calculate the exact failure probability after repeated CNOT operations, and give a smaller upper bound for operation number.

This paper is arranged as follows: in Section \ref{a1}, we describe the model of cold ion-trap QC, upon which further discussions are based. In Section \ref{a2} we obtain the state of the qubits after one CNOT operation. In Section \ref{a3} failure probability after many CNOT operations is calculated. In Section \ref{a4} we give out some discussions. In Section \ref{a5} some conclusions are reached.

\section{Modeling}
\subsection{CNOT gate in Cirac-Zoller scheme}

We consider the implementation of CNOT gate in Cirac-Zoller scheme. There are five steps in
one CNOT gate, in each step, a $k\pi$ pulse is used,
whose duration satisfies $gt_0\sqrt{\bar{n}}=k\pi/2$
\cite{Enk02}, here $\bar{n}$ is mean number of photons in the
pulse.

For each ion, the two levels of the qubit are denoted as
$|0\rangle$ and $|1\rangle$; an auxiliary level $|aux\rangle$
is also needed when implementing CNOT gate. Suppose $x$th ion
is the control qubit, $y$th ion is the target qubit, and the
phonon of center-of-mass mode is the bus qubit. Denote $\omega_0$ the frequency difference between $|0\rangle$ and $|1\rangle$, and $\omega_t$ the phonon frequency, then the five steps are \cite{Cirac&Zoller1995,Steane1997}:

(1) Apply a $\pi/2$ pulse to the $y$th ion, the frequency $\omega$ satisfies $\omega=\omega_0$, and the laser phase is $\phi=-\pi/2$;

(2) Apply a $\pi$ pulse to the $x$th ion, $\omega=\omega_0-\omega_t$, and $\phi=0$;

(3) Apply a $2\pi$ pulse to the $y$th ion,
$\omega=\omega'-\omega_t$, here $\omega'$ is the frequency
difference between the ground state $|0\rangle$ and the
auxiliary state $|aux\rangle$, and $\phi=0$;

(4) The same as step (2);

(5) Apply a $\pi/2$ pulse to the $y$th ion, $\omega=\omega_0$, and $\phi=\pi/2$.

Then we can see that for each step
of one CNOT gate, the system is comprised of a two-level ion,
motional degree of freedom (phonon) and a laser field (in free space), and the
interaction time is within one Rabi period.

\subsection{The model}

The model in \cite{Luo1998} gives an all-quantized
Jaynes-Cummings model(JCM) \cite{Jaynes&Cummings1963} for a cold trapped ion. The system
they consider consists of a two-level trapped ion satisfying
the Lamb-Dicke limit and the strong confinement
limit\cite{Monroe95, Blockley92, Wu97}, and a single-mode
quantized radiation field. The term ``all-quantized" means that
not only the motional degree of the ion is quantized, but also
the radiation field is quantized.

The system we are dealing with is a little different from that
of \cite{Luo1998}: the laser field in our case is in the free
space, thus is a multi-mode field. The laser field in \cite{Luo1998} is in cavity, then only contains one mode. According to \cite{Yang12}, for the purpose of estimating the upper bound of CNOT operation number, it is valid to use JCM for free-space case. Then the model in \cite{Luo1998} can be applied to the problem here.

For each step of Cirac-Zoller scheme, the
Hamiltonian in the interaction picture is
\begin{eqnarray}
H&=&\frac{1}{2}\Omega\sin\theta(a^\dagger\sigma_-\rm e^{\rm i(\omega-\omega_0)t}\rm e^{-\rm i\phi}+a^\dagger\sigma_+\rm e^{\rm i(\omega+\omega_0)t}\rm e^{\rm i\phi}+h.c.)\nonumber \\
&&+\frac{1}{2}\eta\Omega\cos\theta(a^\dagger b^\dagger\sigma_-\rm e^{\rm i(\omega-\omega_0+\omega_t)t}\rm e^{-\rm i\phi}+a^\dagger b\sigma_-\rm e^{\rm i(\omega-\omega_0-\omega_t)t}\rm e^{-\rm i\phi}\nonumber \\
&&+a^\dagger b^\dagger\sigma_+\rm e^{\rm i(\omega+\omega_0+\omega_t)t}\rm e^{\rm i\phi}+a^\dagger b\sigma_+\rm e^{\rm i(\omega+\omega_0-\omega_t)t}\rm e^{\rm i\phi}+h.c.) \nonumber,
\end{eqnarray}
where $\Omega$ is the coupling parameter between ion and radiation, $\theta$ accounts for the relative position of the CM of the ion to the standing wave, $a^\dagger$,$a$ are the creation and destruction operators of the radiation field, $b^\dagger$,$b$ are the creation and destruction operators of the CM vibrational phonons, $\sigma_+$ and $\sigma_-$ are the raising and lowering operators of the two-level ion. Taking the rotating-wave approximation, there are two important cases that we are interested in:
1. Carrier excitation with $\omega=\omega_0$, with
\begin{eqnarray*}
H_1=\hbar g(a^\dagger\sigma_-{\rm e}^{-\rm i\phi}+ a \sigma_+\rm e^{\rm i\phi}),
\end{eqnarray*}
where $g=\frac{1}{2}\Omega\sin\theta$, the corresponding unitary operator is
\begin{small}
\begin{eqnarray}\small \label{e1}
U_1(t)&=&\cos \left(gt\sqrt{a^\dag a+1}\right)|1\rangle\langle1|+\cos \left(gt\sqrt{a^\dag a}\right)|0\rangle\langle0|\nonumber\\
&&-{\rm i}\left[{\rm e}^{{\rm i}\phi}\frac{\sin \left(gt\sqrt{a^\dag a+1}\right)}{\sqrt{a^\dag a+1}}a|1\rangle\langle0|+{\rm e}^{-{\rm i}\phi}a^\dag\frac{\sin\left (gt\sqrt{a^\dag a+1}\right)}{\sqrt{a^\dag a+1}}|0\rangle\langle1|\right].
\end{eqnarray}
\end{small}
If the initial
state is $|\varphi(0)\rangle=|0\rangle_{\rm ion} |n\rangle_{\rm l}$ (ion
represents atom, and l represents laser), then
\[
|\varphi(t)\rangle=U_1 (t)|\varphi(0)\rangle=\cos gt\sqrt{n}|0\rangle_{\rm ion} |n\rangle_{\rm l}-{\rm i}{\rm e}^{{\rm i}\phi}\sin gt \sqrt{n}|1\rangle_{\rm ion} |n-1\rangle_{\rm l}.
\]
Similarly, if the initial state is
$|\varphi(0)\rangle=|1\rangle_{\rm ion}|n\rangle_{\rm l}$, then
\[
|\varphi(t)\rangle=U _1(t)|\varphi(0)\rangle=\cos gt\sqrt{n+1}|1\rangle_{\rm ion} |n\rangle_{\rm l}-{\rm i}{\rm e}^{{-\rm i}\phi}\sin gt \sqrt{n+1}|0\rangle_{\rm ion} |n+1\rangle_{\rm l}.
\]

2. Red sideband excitation with
$\omega=\omega_0-\omega_t$, then
\begin{eqnarray*}
H_2=g'(a^\dagger b^\dagger\sigma_-{\rm e}^{-{\rm i}\phi}+ab\sigma_+{\rm e}^{{\rm i}\phi})
\end{eqnarray*}
where $g'=\frac{1}{2}\eta\Omega\cos\theta$, the unitary operator is
{

\begin{eqnarray} \label{e2}
U_2 (t)&=&\cos \left(g't\sqrt{(a^\dag a+1)(b^\dag b+1)}\right)|1\rangle\langle1|+\cos \left(g't\sqrt{a^\dag ab^\dag b}\right)|0\rangle\langle0|\nonumber\\
&&-{\rm i}[{\rm e}^{{\rm i}\phi}\frac{\sin \left(g't\sqrt{(a^\dag a+1)(b^\dag b+1)}\right)}{\sqrt{(a^\dag a+1)(b^\dag b+1)}}ab|1\rangle\langle0|\nonumber\\
&&+{\rm e}^{-{\rm i}\phi}a^\dag b^\dag\frac{\sin\left (g't\sqrt{(a^\dag a+1)(b^\dag b+1)}\right)}{\sqrt{(a^\dag a+1)(b^\dag b+1)}}|0\rangle\langle1|],
\end{eqnarray}

}
For example, if the initial state is
$|\varphi(0)\rangle=|1\rangle_{\rm ion} |0\rangle_{\rm ph}|n\rangle_{\rm l}$
(ph represents phonons), then
\begin{eqnarray*}
|\varphi(t)\rangle&=&U _2(t)|\varphi(0)\rangle=\cos g't\sqrt{n+1}|1\rangle_{\rm ion}
|0\rangle_{\rm ph}|n\rangle_{\rm l}\nonumber\\
&&-{\rm i}{\rm e}^{-{\rm i}\phi}\sin g't \sqrt{n+1}|0\rangle_{\rm ion}
|1\rangle_{\rm ph}|n+1\rangle_{\rm l}.
\end{eqnarray*}

\section{State of the qubits after one CNOT gate} \label{a2}

For the $x$th ion, the two-level system we choose as the qubit
has levels $|0\rangle_x$ and $|1\rangle_x$; for the $y$th ion,
in addition to the two-level system containing $|0\rangle_y$
and $|1\rangle_y$, it also has an auxiliary level
$|aux\rangle_y$ to complete CNOT operation. For the phonon, the
two levels used as qubit are denoted as $|0\rangle_{ph}$ and
$|1\rangle_{ph}$. Then for the system of $x$th ion, $y$th ion
and phonon, the initial state is generally superposition of
the 12 computational basis:
\begin{eqnarray}  \label{e4}
|\psi(0)\rangle&&=\alpha_1|0\rangle_x |0\rangle_y |0\rangle_{ph}+\alpha_2|1\rangle_x |0\rangle_y |0\rangle_{ph}\nonumber\\
&&+\alpha_3|0\rangle_x |1\rangle_y |0\rangle_{ph}+\alpha_4|1\rangle_x |1\rangle_y |0\rangle_{ph}\nonumber\\
&&+\alpha_5|0\rangle_x |aux\rangle_y |0\rangle_{ph}+\alpha_6|1\rangle_x |aux\rangle_y |0\rangle_{ph}\nonumber\\
&&+\alpha_7|0\rangle_x |0\rangle_y |1\rangle_{ph}+\alpha_8|1\rangle_x |0\rangle_y |1\rangle_{ph}\nonumber\\
&&+\alpha_9|0\rangle_x |1\rangle_y |1\rangle_{ph}+\alpha_{10}|1\rangle_x |1\rangle_y |1\rangle_{ph}\nonumber\\
&&+\alpha_{11}|0\rangle_x |aux\rangle_y |1\rangle_{ph}+\alpha_{12}|1\rangle_x |aux\rangle_y |1\rangle_{ph}.
\end{eqnarray}
Generally, the phonons are cooled to $|0\rangle$ initially in experiment, and the state of ions is relative simple, then the coefficient of some terms in Eq.~(\ref{e4}) may be zero. However, for the purpose of getting the quantum transformation of the qubits after one CNOT gate, it is necessary to let the initial state to be the general state in Eq.~(\ref{e4}), so as to obtain the right transformation.

The single-mode quantized coherent field can be written as
$|\psi^{(n)}\rangle_l=\sum_{n=0}^{\infty }c_{{n}}|n\rangle_l$,
here $|c_n|^2=\frac{{\rm e}^{-\bar{n}}\bar{n}^n}{n!}$. This expression is summed over $n$, then it seems unnecessary to label the state with $n$. However, because the laser fields are different in each step of one CNOT gate, the parameter $n$ can be seen as a way to distinguish the fields. Then for
the first step of CNOT gate, the initial state can be written
as $|\psi'(0)\rangle=|\psi(0)\rangle\otimes
|\psi^{(m)}\rangle_l$.

\begin{bfseries}
\flushleft{[Step 1]}
\end{bfseries}
The interaction is carrier excitation, then the unitary
evolution operator $U_1(t)$ in Eq.~(\ref{e1}) is applicable.
The laser field and the $y$th ion are evolved. The duration
$t_1$ of $\pi/2$ pulse satisfies $gt_1\sqrt{\bar{m}}=\pi/4$,
and $\phi=-\pi/2$. We first obtain
\begin{eqnarray*}
U_1(t_1)\Big(|\psi^{(m)}\rangle_l\otimes|0\rangle_y\Big)&&=\sum_{m=0}^{\infty
}c_{{m}}\cos\frac{\pi\sqrt{m}}{4\sqrt{\bar{m}}}|m\rangle|0\rangle_y\nonumber\\
&&-\sum_{m=0}^{\infty
}c_{{m}}\sin\frac{\pi\sqrt{m}}{4\sqrt{\bar{m}}}|m-1\rangle|1\rangle_y\nonumber\\
&&\stackrel{\triangle}{=}A_1(m)|0\rangle_y-A_2(m)|1\rangle_y,
\end{eqnarray*}
and
\begin{eqnarray*}
U_1(t_1)\Big(|\psi^{(m)}\rangle_l\otimes|1\rangle_y\Big)&&=\sum_{m=0}^{\infty
}c_{{m}}\cos\frac{\pi\sqrt{m+1}}{4\sqrt{\bar{m}}}|m\rangle|1\rangle_y\nonumber\\
&&+\sum_{m=0}^{\infty
}c_{{m}}\sin\frac{\pi\sqrt{m+1}}{4\sqrt{\bar{m}}}|m+1\rangle|0\rangle_y\nonumber\\
&&\stackrel{\triangle}{=}A_3(m)|1\rangle_y-A_4(m)|0\rangle_y.
\end{eqnarray*}
Let the state of the whole system ($x$th ion, $y$th ion and the phonon) after step 1 be $|\psi^{(1)}_a\rangle$, then we can obtain the expression of $|\psi^{(1)}_a\rangle$, because it is lengthy, it is given in Appendix \ref{a6}. (The treatment to the expressions of states after step 2 to step 5 is similar.)

\begin{bfseries}
\flushleft{[Step 2]}
\end{bfseries}

The interaction is red sideband excitation, then the unitary
evolution operator $U_2(t)$ in Eq.~(\ref{e2}) is applicable.
A second laser field, the $x$th ion and the phonons are evolved. For
this step, the initial state can be written as
$|\psi(t_1)\rangle\otimes |\psi^{(n)}\rangle_l$. The duration
$t_2$ of $\pi$ pulse satisfies $gt_2\sqrt{\bar{n}}=\pi/2$, and
$\phi=0$. From
\begin{eqnarray*}
U_2(t_2)\Big(|\psi^{(n)}\rangle_l|1\rangle_x|0\rangle_{ph}\Big)&&=\sum_{n=0}^{\infty
}c_{{n}}\cos\frac{\pi\sqrt{n+1}}{2\sqrt{\bar{n}}}|n\rangle|1\rangle_x|0\rangle_{ph}\nonumber\\
&&-{\rm i}\sum_{n=0}^{\infty
}c_{{n}}\sin\frac{\pi\sqrt{n+1}}{2\sqrt{\bar{n}}}|n+1\rangle|0\rangle_x|1\rangle_{ph}\nonumber\\
&& \stackrel{\triangle}{=}A_5(n)|1\rangle_x|0\rangle_{ph}-{\rm i}A_6(n)|0\rangle_x|1\rangle_{ph},
\end{eqnarray*}
and
\begin{eqnarray*}
U_2(t_2)\Big(|\psi^{(n)}\rangle_l|0\rangle_x|1\rangle_{ph}\Big)&&=\sum_{n=0}^{\infty
}c_{{n}}\cos\frac{\pi\sqrt{n}}{2\sqrt{\bar{n}}}|n\rangle|0\rangle_x|1\rangle_{ph}\nonumber\\
&&-{\rm i}\sum_{n=0}^{\infty
}c_{{n}}\sin\frac{\pi\sqrt{n}}{2\sqrt{\bar{n}}}|n-1\rangle|1\rangle_x|0\rangle_{ph}\nonumber\\
&& \stackrel{\triangle}{=}A_7(n)|0\rangle_x|1\rangle_{ph}-{\rm i}A_8(n)|1\rangle_x|0\rangle_{ph},
\end{eqnarray*}
we can obtain the state of the whole system after step 2 $|\psi^{(1)}_b\rangle$.

\begin{bfseries}
\flushleft{[Step 3]}
\end{bfseries}
The interaction is red sideband excitation, then the unitary
evolution operator is $U_2(t)$ in Eq.~(\ref{e2}). A third laser
field, the $y$th ion and the phonons are evolved. For this
step, the initial state can be written as
$|\psi(t_1+t_2)\rangle\otimes |\psi^{(p)}\rangle_l$. The
duration $t_3$ of $2\pi$ pulse satisfies
$gt_3\sqrt{\bar{p}}=\pi$, and $\phi=0$. From
\begin{eqnarray*}
U_2(t_3)\Big(|\psi^{(n)}\rangle_l|0\rangle_y|1\rangle_{ph}\Big)&&=\sum_{p=0}^{\infty
}c_{{p}}\cos\frac{\pi\sqrt{p+1}}{\sqrt{\bar{p}}}|p\rangle|0\rangle_y|1\rangle_{ph}\nonumber\\
&&-{\rm i}\sum_{p=0}^{\infty
}c_{{p}}\sin\frac{\pi\sqrt{p+1}}{\sqrt{\bar{p}}}|p+1\rangle|aux\rangle_y|0\rangle_{ph}\nonumber\\
&& \stackrel{\triangle}{=}A_9(p)|0\rangle_y|1\rangle_{ph}-{\rm i}A_{10}(p)|aux\rangle_y|0\rangle_{ph},
\end{eqnarray*}
and
\begin{eqnarray*}
U_2(t_3)\Big(|\psi^{(p)}\rangle_l|aux\rangle_y|0\rangle_{ph}\Big)&&=\sum_{p=0}^{\infty
}c_{{p}}\cos\frac{\pi\sqrt{p}}{\sqrt{\bar{p}}}|p\rangle|aux\rangle_y|0\rangle_{ph}\nonumber\\
&&-{\rm i}\sum_{p=0}^{\infty
}c_{{p}}\sin\frac{\pi\sqrt{p}}{\sqrt{\bar{n}}}|p-1\rangle|0\rangle_y|1\rangle_{ph}\nonumber\\
&& \stackrel{\triangle}{=}A_{11}(p)|aux\rangle_y|0\rangle_{ph}-{\rm i}A_{12}(p)|0\rangle_y|1\rangle_{ph},
\end{eqnarray*}
we obtain the state of the whole system after step 3 $|\psi^{(1)}_c\rangle$.

\begin{bfseries}
\flushleft{[Step 4]}
\end{bfseries}

This step is the same to step 2. Assume the state of the field is $|\psi^{(q)}\rangle_l=\sum_{q=0}^{\infty }c_{{q}}|q\rangle_l$, then we can obtain the state of the whole system after step 4 $|\psi^{(1)}_d\rangle$.

\begin{bfseries}
\flushleft{[Step 5]}
\end{bfseries}

This step is similar to step 1. Assume the state of the field is $|\psi^{(r)}\rangle_l=\sum_{r=0}^{\infty }c_{{r}}|r\rangle_l$, then we can obtain the state of the whole system after the step 5 $|\psi^{(1)}_e\rangle\stackrel{\triangle}{=}|\psi^{(1)} \rangle$.
(The expression of $|\psi^{(1)}_i\rangle (i=a, b, c, d, e)$ can be seen in Appendix \ref{a6}.)

Then the density matrix after one CNOT gate is
${\rho}_{total}^{(1)}=|\psi^{(1)} \rangle\langle\psi^{(1)} |$.
This matrix contains the information of the qubits (xth ion, yth ion
and phonon) and the field, but we are interested only in the
qubits. Thus we trace out the five fields and obtain the reduced density matrix $\rho^{(1)}$, which
describe the state of $x$th ion, $y$th ion and phonon. It has
144 elements, the simplest elements is:

\begin{equation*}
\begin{split}
|\rho^{(1)}\rangle_{11}&=\alpha_1^* \Big[\alpha_1S_2^*S_4^*-\alpha_3S_3^*S_4^*+ \alpha_3S_4^*S_1^*+\alpha_1S_1S_1\\
&-\alpha_3S_6^*S_{10}+\alpha_3S_7^*S_4+\alpha_1S_2S_4+
\alpha_1S_5S_{10}\Big]\\
&+\alpha_3^* \Big[\alpha_3S_{10}S_1+\alpha_1S_4S_1- \alpha_3S_9^*S_4-\alpha_1S_3S_4\\
&-\alpha_1S_6S_{10}+S_4^*(-\alpha_3S_9+\alpha_1S_7)+
\alpha_3S_8S_{10}\Big],
\end{split}
\end{equation*}

here
\begin{equation*}
\begin{split}
&S_1=\sum _{n=0}^{\infty }\frac{{{\rm e}^{-\bar{n}}}\bar{n}^n}{n!}\cos^2 (\frac{\pi\sqrt{n}}{4\sqrt{\bar{n}}}),\\
&S_2=\sum _{n=0}^{\infty }\frac{{{\rm e}^{-\bar{n}}}\bar{n}^n}{n!}\sqrt{\frac{\bar{n}}{n+1}}\cos (\frac{\pi\sqrt{n}}
{4\sqrt{\bar{n}}})\sin (\frac{\pi\sqrt{n+1}}{4\sqrt{\bar{n}}}),\\
&S_3=\sum _{n=0}^{\infty }\frac{{{\rm e}^{-\bar{n}}}\bar{n}^n}{n!}\cos (\frac{\pi\sqrt{n}}{4\sqrt{\bar{n}}})
\cos (\frac{\pi\sqrt{n+1}}{4\sqrt{\bar{n}}}),\\
&S_4=\frac{1}{2}\sum _{n=0}^{\infty }\frac{{{\rm e}^{-\bar{n}}}\bar{n}^n}{n!}\sqrt{\frac{n}{\bar{n}}}\sin (\frac{k\pi\sqrt{n}}{2\sqrt{\bar{n}}}),\\
&S_5=\sum _{n=0}^{\infty }\frac{{{\rm e}^{-\bar{n}}}\bar{n}^n}{n!}\sin^2 (\frac{\pi\sqrt{n}}{4\sqrt{\bar{n}}}),
\end{split}
\end{equation*}
\begin{equation*}
\begin{split}
&S_6=\frac{1}{2}\sum _{n=0}^{\infty }\frac{{{\rm e}^{-\bar{n}}}\bar{n}^n}{n!}\sqrt{\frac{\bar{n}}{n+1}}\sin (\frac{\pi\sqrt{n+1}}{2\sqrt{\bar{n}}}),\\
&S_7=\sum _{n=0}^{\infty }\frac{{{\rm e}^{-\bar{n}}}\bar{n}^n}{n!}\sqrt{\frac{n}{n+1}}\sin (\frac{\pi\sqrt{n}}
{4\sqrt{\bar{n}}})\sin (\frac{\pi\sqrt{n+1}}{4\sqrt{\bar{n}}}),\\
&S_8=\sum _{n=0}^{\infty }\frac{{{\rm e}^{-\bar{n}}}\bar{n}^n}{n!}\cos^2 (\frac{\pi\sqrt{n+1}}{4\sqrt{\bar{n}}}),\\
&S_9=\sum _{n=0}^{\infty }\frac{{{\rm e}^{-\bar{n}}}\bar{n}^n}{n!}\sqrt{\frac{n}{\bar{n}}}\cos (\frac{\pi\sqrt{n+1}}{4\sqrt{\bar{n}}})\sin (\frac{\pi\sqrt{n}}{4\sqrt{\bar{n}}}),\\
&S_{10}=\sum _{n=0}^{\infty }\frac{{{\rm e}^{-\bar{n}}}\bar{n}^n}{n!}\sin^2 (\frac{\pi\sqrt{n+1}}{4\sqrt{\bar{n}}}).
\end{split}
\end{equation*}
There are 42 sums like $S_i(i=1,\cdots,10)$ above in $\rho^{(1)}$, and all these sums can be calculated
accurately using the algorithm in \cite{Yang12}, which can achieve any given precision. The value of
$S_i~(i=1,\cdots,10)$ are shown in Table.~\ref{t1}, here $\bar{n}=10^4$, $k=2$.
\begin{table}[htbp]
\caption{\label{t1}  Values of $S_i (i=1,2,\cdots,10)$} \centering
\begin{tabularx}{240pt}{p{30pt}p{210pt}}
\hline\hline
 \textrm{Sum}&
\textrm{~~~~~~~~~~~~~~~~~~~~Value~~~~~~~~~~~~~~~~~~~~}\\
 \hline
$S_1$&$0.500009817401897928264355667147$\\
$S_2$&$0.499997963670545683314749144417$\\
$S_3$&$0.499990182422436977119909877501$\\
$S_4$&$0.499978328996648238077753901338$\\
$S_5$&$0.499990182598102071735644332853$\\
$S_6$&$0.499978328996648238077753901338$\\
$S_7$&$0.499984817654121934189482133164$\\
$S_8$&$0.499970548214032003626335555500$\\
$S_9$&$0.499958694533432311848321647856$\\
$S_{10}$&$0.500029451785967996373664444500$\\
\hline\hline
\end{tabularx}
\end{table}
\section{Failure probability after many CNOT operations} \label{a3}

\subsection{Quantum transformation after one CNOT operation}

Consider the state transformation for the qubits after one CNOT
operation with initial state
$\rho^{(0)}=|\psi^{(0)}\rangle\langle\psi^{(0)}|$, here $|\psi^{(0)}\rangle$ is defined in Eq.~(\ref{e4}). For step 1 of CNOT gate, the
operator-sum representation of hyperoperator $\mathcal{E}$ in quantum open
system has the result $
\rho_a^{(1)}=\mathcal{E}(\rho^{(0)})=\sum_iE_i\rho^{(0)}E_i^\dag.$
If we let
\[M_1=E_1\otimes (E_1^\dag)^T+E_2\otimes (E_2^\dag)^T+\cdots=E_1\otimes E_1^*+E_2\otimes E_2^*+\cdots,
\]
then $\vec{\rho}_a^{(1)}=M_1\vec{\rho}^{\,(0)}$, here
$\vec{\rho}$ is straightening of matrix $\rho$. Similarly, if
we let the corresponding matrix of step 2 to step 5 be
$M_2$,$M_3$,$M_4$,$M_5$, then
\begin{equation} \label{e3}
\vec{\rho}^{(1)}=M\vec{\rho}^{\,(0)},
\end{equation}
here
$M=M_5M_4M_3M_2M_1$.

\subsection{Calculation of $M$}
According to the previous calculation, $\rho^{(0)}$,
$\rho^{(1)}$ are both $12\times 12$ matrices. Then
$\vec{\rho}^{(0)}$, $\vec{\rho}^{(1)}$ are $144\times 1$ column
vectors. From Eq.~(\ref{e3}), we can see that the matrix $M$
related to one CNOT operation is of the size $144\times
144$. We need to determine the $144\times 144$ elements.

From Eq.~(\ref{e3}), we can only get 144 equations from the rule of matrix multiplication. Examine
each equation, because $\alpha_1,\cdots,\alpha_{12}$ is any set
of complex numbers with modular square sum 1, then the
coefficient of $\alpha_1\alpha_1^*,\alpha_1\alpha_2^*$,\\$\cdots,
\alpha_{12}\alpha_{12}^*$ should be the same on both sides of each
equation. Finally we have $144\times 144$ equations in all, which
is sufficient to determine $M$. We find that $M$ has no relation with $\alpha_i(i=1,\cdots,12)$, thus is independent of initial state.
\subsection{Final state and failure probability}
Once $M$ is obtained, from
$\vec{\rho}^{(1)}=M\vec{\rho}^{\,(0)}$, we can get the state
after $t$ times of CNOT operation
$\vec{\rho}^{(t)}=M\vec{\rho}^{\,(t-1)}=\cdots=M^t\vec{\rho}^{\,(0)}$,
then ${\rho}^{(t)}$ is obtained.

In Cirac-Zoller's CNOT scheme, the phonons are cooled to state
$|0\rangle_{ph}$ in the beginning. The $x$th ion and $y$th ion
are generally in the superposition of computational basis
$\{|0\rangle_x|0\rangle_y,
|0\rangle_x|1\rangle_y,|1\rangle_x|0\rangle_y,|1\rangle_x|1\rangle_y\}$.
Thus the actual initial state of CNOT operation is
$|\tilde{\psi^{(0)}}\rangle=\alpha_1|0\rangle_x |0\rangle_y
|0\rangle_{ph}+\alpha_2|1\rangle_x |0\rangle_y
|0\rangle_{ph}+\alpha_3|0\rangle_x |1\rangle_y
|0\rangle_{ph}+\alpha_4|1\rangle_x |1\rangle_y |0\rangle_{ph}$.
For the final state ${\rho}^{(t)}$, we are interested in the
state of ions, then we can trace the freedom of phonons out to
get $\rho^{(t)'}$. The actual final state is
$\widetilde{\rho^{(t)}}=\rho^{(t)'}|_{\alpha_5,\cdots,\alpha_{12}=0}$.
To obtain the failure probability of CNOT operation, we first need to know the expected state. It can be written as
$|\psi_e\rangle=\alpha_1|0\rangle_x |0\rangle_y
+\alpha_2|1\rangle_x |0\rangle_y +\alpha_3|0\rangle_x
|1\rangle_y +\alpha_4|1\rangle_x |1\rangle_y $ when the number
of CNOT operations is even. When the operation number is odd,
the expected state is $|\psi_e\rangle=\alpha_1|0\rangle_x
|0\rangle_y +\alpha_4|1\rangle_x |0\rangle_y
+\alpha_3|0\rangle_x |1\rangle_y +\alpha_2|1\rangle_x
|1\rangle_y $. The success probability is
\[
p_s=\langle\psi_e|\tilde{\rho^{(t)}}|\psi_e\rangle,
\]
then the failure probability is $p_f=1-p_s$.

\begin{figure}[htbp]
\raggedright
\subfloat[]{
\begin{minipage}[b]{0.5\textwidth}
\centering

\includegraphics[width=2.7in,height=1.5in]{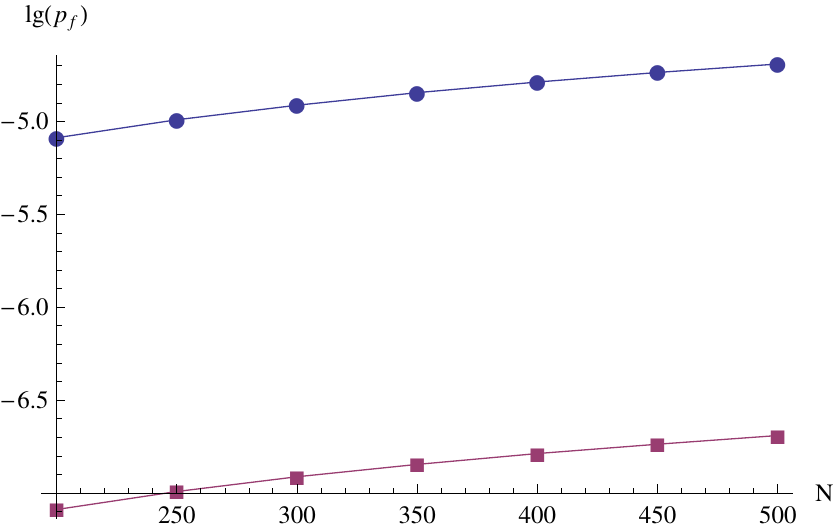}
\end{minipage}
}
\subfloat[]{
\begin{minipage}[b]{0.5\textwidth}
\centering

\includegraphics[width=2.7in,height=1.5in]{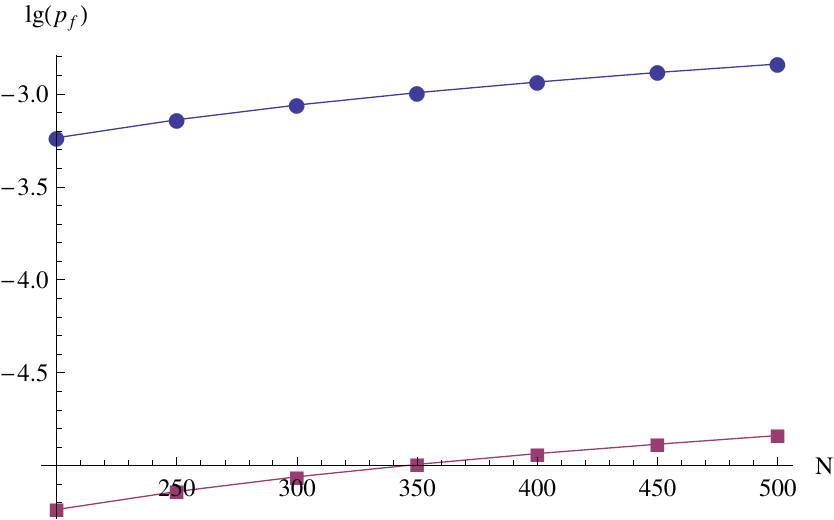}
\end{minipage}
}

\subfloat[]{
\begin{minipage}[b]{0.5\textwidth}
\centering

\includegraphics[width=2.7in,height=1.5in]{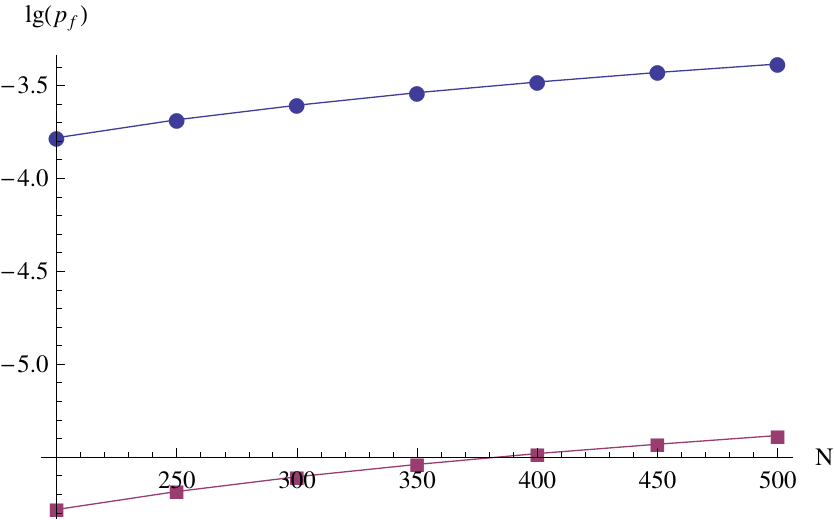}
\end{minipage}
}
\subfloat[]{
\begin{minipage}[b]{0.5\textwidth}
\centering

\includegraphics[width=2.7in,height=1.5in]{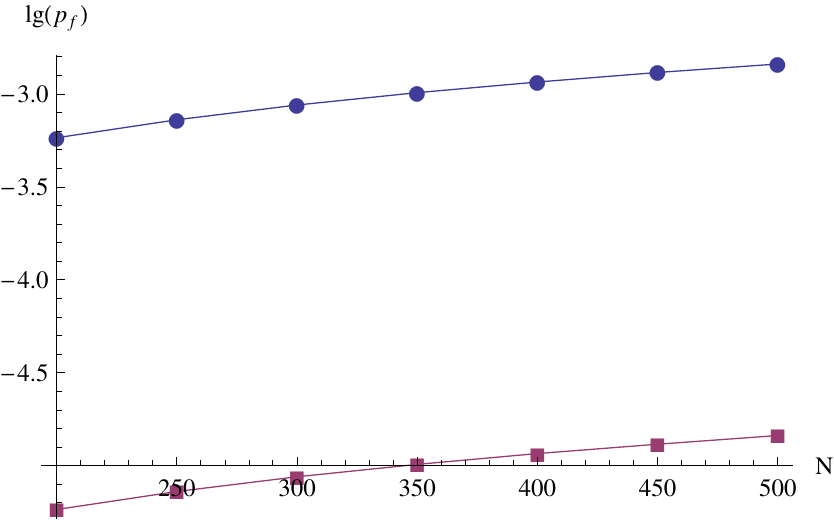}
\end{minipage}
}

\subfloat[]{
\begin{minipage}[b]{0.5\textwidth}
\centering

\includegraphics[width=2.7in,height=1.5in]{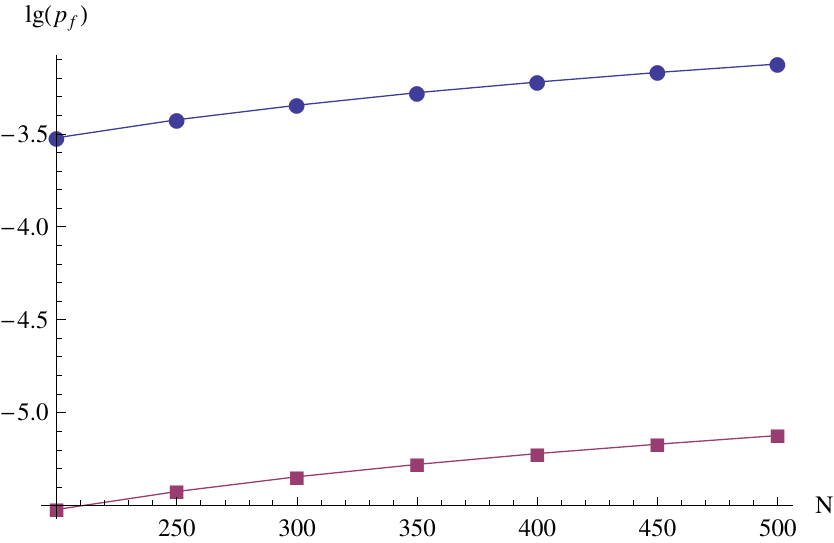}
\end{minipage}
}
\subfloat[]{
\begin{minipage}[b]{0.5\textwidth}
\centering
\includegraphics[width=2.7in,height=1.5in]{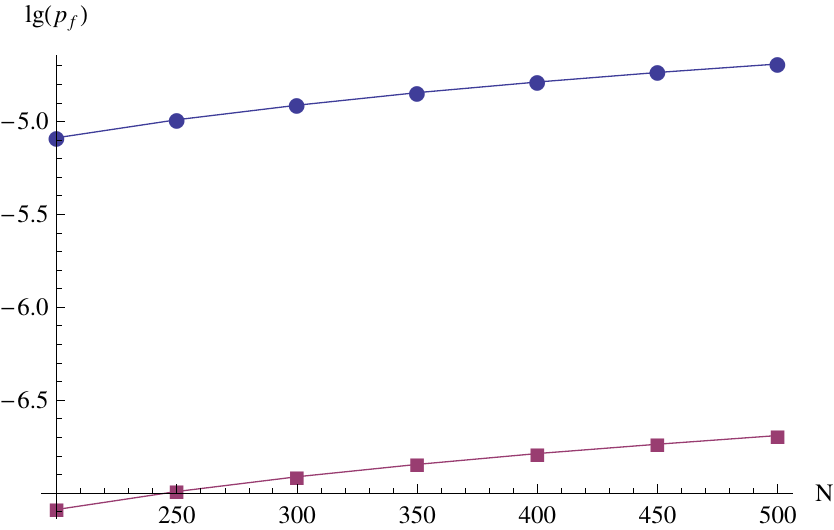}
\end{minipage}
} \caption{Failure probability from field quantization in cases of different initial state and $\bar{n}$}\label{ff1}
\end{figure}

The logarithm of failure
probability to operation number in the cases of different initial state and mean number of photons are plotted in
Fig.~\ref{ff1}. In each figure, the curve above represents $\bar{n}=10^6$ and the curve above represents $\bar{n}=10^8$. The initial states in (a) to (e) are $|0\rangle_x
|0\rangle_y$,$|1\rangle_x |0\rangle_y$,$|0\rangle_x
|1\rangle_y$,$|1\rangle_x
|1\rangle_y$,
$1/\sqrt{2}(|0\rangle_x |0\rangle_y+|1\rangle_x
|0\rangle_y)$,$1/\sqrt{2}(|0\rangle_x |0\rangle_y+|0\rangle_x
|1\rangle_y)$, respectively.

Then we can see that failure
probability is proportional to operation number, and inverse
proportional to mean number of photons. It is different for different initial states: if the initial state is $|1\rangle_x
|0\rangle_y$ or $|1\rangle_x |1\rangle_y$, the failure probability is relatively large; if the initial state is $|0\rangle_x |0\rangle_y$ or $|0\rangle_x |1\rangle_y$, the failure probability is relatively small. If the initial state is a superposition of the four states above, then the same rule works:  if the amplitude of $|1\rangle_x
|0\rangle_y$ or $|1\rangle_x |1\rangle_y$ is large, then the
failure probability is large; however, if the amplitude of
$|0\rangle_x |0\rangle_y$ or $|0\rangle_x |1\rangle_y$ is
large, then the failure probability is small.

\subsection{Failure probability when the limitation to $\bar{n}$ in sideband transition is considered}

For the failure probability obtained in Sec.~\ref{a3}, we
cannot decrease the error by simply increasing mean number of
photons. As is known, a large mean number of photons means a
more intense laser. Then there is a limitation in mean number of photons in practical realization of quantum computation. A strong field may force the electron in the ion out of the nucleus' bounding, then destroy the quantum computation system. Besides, the devices may not work after the strong laser goes through it. Our former work shows that for sideband transition, the mean number of photons cannot be more than $10^4$ when the wavelength is of the order $10^{-6}$ m \cite{Yang12}.

From Sec.~\ref{a2}, the transition in step 2 to step 4 are sideband transition. Then we assume step 1 and step 5 are accurately implemented, and take into account the limitation to mean number of photons in step 2 to step 4. Using the method in Sec.~\ref{a3} and \ref{a4}, we get the lower bound of failure probability. The
result is when $\bar{n}=10^4$ and the initial state is
$|1\rangle_x |0\rangle_y$ or
 $|1\rangle_x |1\rangle_y$, after $10^2$ times of
CNOT operations, failure probability reaches $10^{-2}$. The
failure probability under other initial states is relatively
smaller, but no less than $10^{-4}$.
\section{Discussion} \label{a4}

\subsection{The permitted depth of CNOT operations in Cirac-Zoller scheme}

The threshold theorem in quantum computation declares that an
arbitrarily long quantum computation can be performed reliably
if the failure probability of each quantum gate is less than a
critical value. The recognized maximum threshold is $10^{-2}$ \cite{Knill2005}. Then when $\lambda=10^{-6}$~m,and
the initial state is $|1\rangle_x |0\rangle_y$ or
 $|1\rangle_x |1\rangle_y$, the number
of CNOT gates on the same pair of physical qubits should be no more than $10^2$ in one
error-correction period, or else the computation cannot be
implemented reliably. For other initial states, the number of CNOT operations also has limitations. This result can help to determine the number of CNOT operations on the same pair of physical qubits between coding and decoding in one error-correction period.

\subsection{Difference between this paper and \cite{Yang12}}

\subsection{The reason to calculate the failure probability after repeated CNOT operations}

\section{Conclusion} \label{a5}

Firstly, we obtain the exact expression of failure probability from field quantization after many CNOT operations in Cirac-Zoller scheme. Using the
operator-sum representation of hyperoperator in quantum open
system, we obtain the quantum transformation of the qubits
after one CNOT operation, then the state after many CNOT gates
can be obtained. Comparing the actual result with expected
states under semiclassical treatment, we can obtain the failure
probability after many CNOT operations. The
conclusions we arrive at are: failure probability is proportional to operation number and the amplitude of $|1\rangle_x
|0\rangle_y$ or $|1\rangle_x |1\rangle_y$ in initial state, and inverse
proportional to mean number of photons and amplitude of
$|0\rangle_x |0\rangle_y$ or $|0\rangle_x |1\rangle_y$ in initial state.

Secondly, we calculate the failure probability when the limitation to $\bar{n}$ in sideband transition is considered. Assuming the carrier
transition is accurately carried out, we calculate the lower
bound of the failure probability in the practical case. The
result is when the initial state is
$|1\rangle_x |0\rangle_y$ or
 $|1\rangle_x |1\rangle_y$, after $10^2$ times of
CNOT operations, failure probability is no less than $10^{-2}$. The
failure probability under other initial states is relatively
smaller, but no less than $10^{-4}$. Then when
the initial state is $|1\rangle_x |0\rangle_y$ or
 $|1\rangle_x |1\rangle_y$, the number
of CNOT gates on the same pair of physical qubits should be no more than $10^2$ in one
error-correction period, or else the computation cannot be
implemented reliably. For other initial states, the number of CNOT operations also has limitations.

The conclusions the article arrives at can help determine how
many CNOT operations on the same pair of physical qubits are
permitted during one error-correction period in fault-tolerant
quantum computation under certain initial state. Then this can
be a reference for people designing quantum circuit or quantum
algorithm, which is important to future
study on both theory and experiment.
\appendix

\section{The state after each step in one CNOT gate} \label{a6}

\begin{bfseries}
\flushleft{[State after step 1]}
\end{bfseries}
\begin{eqnarray*}
|\psi^{(1)}_a\rangle&=&(\alpha_1A_1(m)+\alpha_3A_4(m))|0\rangle_x |0\rangle_y|0\rangle_{ph}\nonumber\\
&&+(\alpha_2A_1(m)+\alpha_4A_4(m))|1\rangle_x |0\rangle_y|0\rangle_{ph}\nonumber\\
&&+(-\alpha_1A_2(m)+\alpha_3A_3(m))|0\rangle_x |1\rangle_y|0\rangle_{ph}\nonumber\\
&&+(-\alpha_2A_2(m)+\alpha_4A_3(m))|1\rangle_x |1\rangle_y|0\rangle_{ph}\nonumber\\
&&+\alpha_5|\psi^{(m)}\rangle_l|0\rangle_x |aux\rangle_y|0\rangle_{ph}+\alpha_6|\psi^{(m)}\rangle_l|1\rangle_x |aux\rangle_y|0\rangle_{ph}\nonumber\\
&&+(\alpha_7A_1(m)+\alpha_9A_4(m))|0\rangle_x |0\rangle_y|1\rangle_{ph}\nonumber\\
&&+(\alpha_8A_1(m)+\alpha_{10}A_4(m))|1\rangle_x |0\rangle_y|1\rangle_{ph}\nonumber\\
&&+(-\alpha_7A_2(m)+\alpha_9A_3(m))|0\rangle_x |1\rangle_y|1\rangle_{ph}\nonumber\\
&&+(-\alpha_8A_2(m)+\alpha_{10}A_3(m))|1\rangle_x |1\rangle_y|1\rangle_{ph}\nonumber\\
&&+\alpha_{11}|\psi^{(m)}\rangle_l|0\rangle_x |aux\rangle_y|0\rangle_{ph}+\alpha_{12}|\psi^{(m)}\rangle_l|1\rangle_x |aux\rangle_y|0\rangle_{ph}.
\end{eqnarray*}
\begin{bfseries}
\flushleft{[State after step 2]}
\end{bfseries}
\begin{equation*}
\begin{split}
|\psi^{(1)}_b\rangle&=(\alpha_1A_1(m)+\alpha_3A_4(m))|\psi^{(n)}\rangle_l|0\rangle_x |0\rangle_y |0\rangle_{ph}\\
&+[A_1(m)(\alpha_2A_5(n)-{\rm i}\alpha_7A_8(n))+A_4(m)(\alpha_4A_5(n)\\
&-{\rm i}\alpha_9A_8(n))]|1\rangle_x |0\rangle_y |0\rangle_{ph}\\
&+(-\alpha_1A_2(m)+\alpha_3A_3(m))|\psi^{(n)}\rangle_l|0\rangle_x |1\rangle_y |0\rangle_{ph}\\
&+[A_2(m)(-\alpha_2A_5(n)+{\rm i}\alpha_7A_8(n))\\
&+A_3(m)(\alpha_4A_5(n)-{\rm i}\alpha_9A_8(n))]|1\rangle_x |1\rangle_y |0\rangle_{ph}\allowdisplaybreaks\\
&+\alpha_5|\psi^{(m)}\rangle_l|\psi^{(n)}\rangle_l|0\rangle_x |aux\rangle_y |0\rangle_{ph}\\
&+|\psi^{(m)}\rangle_l(\alpha_6A_5(n)-{\rm i}\alpha_{11}A_8(n))|1\rangle_x |aux\rangle_y |0\rangle_{ph}\\
&+[A_1(m)(-{\rm i}\alpha_2A_6(n)+\alpha_7A_7(n))\\
&+A_4(m)(-{\rm i}\alpha_4A_6(n)+\alpha_9A_7(n))]|0\rangle_x |0\rangle_y |1\rangle_{ph}\\
&+(\alpha_8A_1(m)+\alpha_{10}A_4(m))|\psi^{(n)}\rangle_l|1\rangle_x |0\rangle_y |1\rangle_{ph}\\
&+[A_2(m)({\rm i}\alpha_2A_6(n)-\alpha_7A_7(n))\\
&+A_3(m)(-{\rm i}\alpha_4A_6(n)+\alpha_9A_7(n))]|0\rangle_x |1\rangle_y |1\rangle_{ph}\\
&+(-\alpha_8A_2(m)+\alpha_{10}A_3(m))|\psi^{(n)}\rangle_l|1\rangle_x |1\rangle_y |1\rangle_{ph}\\
&+|\psi^{(m)}\rangle_l(-{\rm i}\alpha_6A_6(n)+\alpha_{11}A_7(n))|0\rangle_x |aux\rangle_y |1\rangle_{ph}\\
&+\alpha_{12}|\psi^{(m)}\rangle_l|\psi^{(n)}\rangle_l|1\rangle_x |aux\rangle_y |1\rangle_{ph}.
\end{split}
\end{equation*}
\begin{bfseries}
\flushleft{[State after step 3]}
\end{bfseries}
\begin{equation*}
\raggedright
\begin{split}
&|\psi^{(1)}_c\rangle=(\alpha_1A_1(m)+\alpha_3A_4(m)) |\psi^{(n)}\rangle_l |\psi^{(p)}\rangle_l|0\rangle_x |0\rangle_y |0\rangle_{ph}\\
&+|\psi^{(p)}\rangle_l \Big[A_1(m) (\alpha_2A_5(n)-{\rm i} \alpha_7A_8(n))\\
&+A_4(m)(\alpha_4A_5(n)-{\rm i}\alpha_9A_8(n))\Big]|1\rangle_x |0\rangle_y |0\rangle_{ph}\\
&+(-\alpha_1A_2(m)+\alpha_3A_3(m)) |\psi^{(n)}\rangle_l |\psi^{(p)}\rangle_l|0\rangle_x |1\rangle_y |0\rangle_{ph}\\
&+|\psi^{(p)}\rangle_l \Big[A_2(m) (-\alpha_2A_5(n)+{\rm i} \alpha_7A_8(n))\\
&+A_3(m)(\alpha_4A_5(n)-{\rm i}\alpha_9A_8(n))\Big]|1\rangle_x |1\rangle_y |0\rangle_{ph}\\
&+\Big[\alpha_5A_{11}(p)|\psi^{(m)}\rangle_l |\psi^{(n)}\rangle_l-A_{10}(p) \Big(A_{1}(m) (\alpha_2A_{6}(n)+{\rm i}\alpha_7A_{7}(n))+A_{4}(m)\times\\
& \times(\alpha _4A_{6}(n)+{\rm i}\alpha_9A_{7}(n))\Big)\Big]|0\rangle_x |aux\rangle_y |0\rangle_{ph}\\
&-{\rm i}\Big[\alpha_8 A_{1}(m) A_{10}(p)|\psi^{(n)}\rangle_l+\alpha_{10}A_{10}(p)A_{4}(m) |\psi^{(n)}\rangle_l+A_{11}(p)|\psi^{(m)}\rangle_l\times\\
& \times ({\rm i}\alpha_6 A_{5}(n)+\alpha_{11}A_{8}(n))\Big]|1\rangle_x |aux\rangle_y |0\rangle_{ph}\\
&-{\rm i}\left\{\alpha_5A_{12}(p)|\psi^{(m)}\rangle_l|\psi^{(n)}\rangle_l+A_9(p)\left[A_1(m)(\alpha_2A_6(n)\right]\right\}
|0\rangle_x |0\rangle_y |1\rangle_{ph}\\
&+[(\alpha_8A_1(m)+\alpha_{10}A_4(m))A_9(p)|\psi^{(n)}\rangle_l\\
&+|\psi^{(m)}\rangle_l(-{\rm i} \alpha_6A_5(n)-\alpha_{11}A_8(n))A_12(p)]|1\rangle_x |0\rangle_y |1\rangle_{ph}\\
&+|\psi^{(p)}\rangle_l \Big[A_2(m) ({\rm i}\alpha_2A_6(n)- \alpha_7A_7(n))\\
&+A_3(m)(-{\rm i}\alpha_4A_6(n)+\alpha_9A_7(n))\Big]|0\rangle_x |1\rangle_y |1\rangle_{ph}\\
&+(-\alpha_8A_2(m)+\alpha_{10}A_3(m)) |\psi^{(n)}\rangle_l |\psi^{(p)}\rangle_l|1\rangle_x |1\rangle_y |1\rangle_{ph}\\
&+|\psi^{(m)}\rangle_l |\psi^{(p)}\rangle_l(-{\rm i}\alpha_6A_6(n)+\alpha_{11}A_7(n))|0\rangle_x |aux\rangle_y |1\rangle_{ph}\\
&+\alpha_{12}|\psi^{(m)}\rangle_l |\psi^{(n)}\rangle_l |\psi^{(p)}\rangle_l|1\rangle_x |aux\rangle_y |1\rangle_{ph}.
\end{split}
\end{equation*}
\begin{bfseries}
\flushleft{[State after step 4]}
\end{bfseries}
\begin{equation*}
\begin{split}
&|\psi^{(1)}_d\rangle=\Big(\alpha_1A_1(m)+\alpha_3A_4(m)\Big)|\psi^{(n)}\rangle_l|\psi^{(p)}\rangle_l|\psi^{(q)}\rangle_l|0\rangle_x |0\rangle_y |0\rangle_{ph}\\
&+\Big\{-\alpha_5A_{12}(p)|\psi^{(m)}\rangle_l|\psi^{(n)}\rangle_lA_8(q)\\
&+A_1(m)\Big[|\psi^{(p)}\rangle_lA_5(q)(\alpha_2A_5(n)-{\rm i}\alpha_7A_8(n))\\
&-A_{9}(p)A_8(q)(\alpha_2A_6(n)+{\rm i}\alpha_7A_7(n))\Big]\\
&+A_4(m)\Big[|\psi^{(p)}\rangle_lA_5(q)(\alpha_2A_5(n)-{\rm i}\alpha_7A_8(n))\\
&-A_{9}(p)A_8(q)(\alpha_4A_6(n)+{\rm i}\alpha_9A_7(n))\Big]\Big\}|1\rangle_x |0\rangle_y |0\rangle_{ph}\\&+\Big(-\alpha_1A_2(m)+\alpha_3A_3(m)\Big)|\psi^{(n)}\rangle_l|\psi^{(p)}\rangle_l|\psi^{(q)}\rangle_l|0\rangle_x |1\rangle_y |0\rangle_{ph}\\
&+|\psi^{(p)}\rangle_l\Big\{A_2(m)\Big[-\alpha_2 A_5(n) A_5(q) + {\rm i} \alpha_7 A_5(q) A_8(n) \\
&+ \alpha_2 A_6(n) A_8(q) + {\rm i} \alpha_7 A_7(n) A_8(q)\Big] \\
&+ A_3(m) \Big[\alpha_4 A_5(n) A_5(q)- {\rm i}\Big(\alpha_9 A_5(q) A_8(n)\\
&-{\rm i}\alpha_4 A_6(n)A_8(q)+ \alpha_9 A_7(n)A_8(q)\Big)\Big]\Big\}|1\rangle_x |1\rangle_y |0\rangle_{ph}\\
&+|\psi^{(q)}\rangle_l \Big\{\alpha_5 A_{11}(p)|\psi^{(m)}\rangle_l|\psi^{(n)}\rangle_l - A_{10}(p) \Big[A_1(m) \Big(\alpha_2 A_6(n) + {\rm i} \alpha_7 A_7(n)\Big) \\
&+A_4(m) (\alpha_4 A_6(n)
+{\rm i}\alpha_9 A_8(n))\Big]\Big\}|0\rangle_x |aux\rangle_y |0\rangle_{ph}\\
&-{\rm i}\Big[\alpha_8 A_1(m)A_{10}(p) |\psi^{(n)}\rangle_l A_5(q) + \alpha_{10} A_{10}(p)A_4(m) |\psi^{(n)}\rangle_lA_5(q)\\ &+|\psi^{(m)}\rangle_l\Big(A_{11}(p) A_5(q)({\rm i}\alpha_6 A_5(n) + \alpha_{11}A_8(n)) \\
&+|\psi^{(p)}\rangle_l(-{\rm i}\alpha_6 A_6(n) + \alpha_{11}A_7(n)) A_{8}(q)\Big)\Big]|1\rangle_x |aux\rangle_y |0\rangle_{ph}\\
&-{\rm i}\Big\{\alpha_5 A_{12}(p) |\psi^{(m)}\rangle_l|\psi^{(n)}\rangle_l A_{7}(q) \\
&+  A_{1}(m) \Big[A_{9}(p)\Big(\alpha_2 A_{6}(n) + {\rm i} \alpha_7 A_{7}(n)\Big) A_{7}(q) +|\psi^{(p)}\rangle_l A_{6}(q)\\
 &(\alpha_2 A_{5}(n) - {\rm i} \alpha_7 A_{8}(n))\Big] + A_{4}(m)\Big[A_{9}(p) (\alpha_4 A_{6}(n) +{\rm i}\alpha_9 A_{7}(n)) A_{7}(q)|\psi^{(p)}\rangle_lA_{6}(q)\\
 &\Big(\alpha_4 A_{5}(n) - {\rm i} \alpha_9A_{8}(n)\Big)\Big]\Big\}|0\rangle_x |0\rangle_y |1\rangle_{ph}\\
 \end{split}
\end{equation*}
\begin{equation*}
\begin{split}
&+|\psi^{(q)}\rangle_l \Big[\alpha_8A_1(m)A_9(p)|\psi^{(n)}\rangle_l + \alpha_{10} A_9(p)A_4(m)|\psi^{(n)}\rangle_l +
   A_{11}(p) |\psi^{(m)}\rangle_l\\
   &\Big(-{\rm i}  \alpha_6 A_5(n) - \alpha_{11}A_8(n)\Big)\Big]|1\rangle_x |0\rangle_y |1\rangle_{ph}\\
&+|\psi^{(p)}\rangle_l \Big[A_2(m)\Big({\rm i}\alpha_2 A_5(n) A_6(q) +
      {\rm i} \alpha_2 A_6(n)A_7(q) \\
      &- \alpha_7 A_7(n)A_7(q)+ \alpha_7 A_6(q) A_8(n)\Big) \\
      &+ A_3(m) \Big(-{\rm i}\alpha_4 A_5(n)A_6(q) -
      {\rm i}\alpha_4A_6(n)A_7(q) \\
      &+ \alpha_9A_7(n) A_7(q) - \alpha_9 A_6(q)A_8(n)\Big)\Big]|0\rangle_x |1\rangle_y |1\rangle_{ph}\\
&+(-\alpha_8A_2(m) + \alpha_{10} A_3(m))|\psi^{(n)}\rangle_l|\psi^{(p)}\rangle_l|\psi^{(q)}\rangle_l|1\rangle_x |1\rangle_y |1\rangle_{ph}\\
&-\alpha_8 A_1(m)A_{10}(p)|\psi^{(n)}\rangle_l A_6(q) - \alpha_{10} A_{10}(p) A_4(m)|\psi^{(n)}\rangle_l A_6(q) -|\psi^{(m)}\rangle_l \Big[|\psi^{(p)}\rangle_l\\
& \Big({\rm i} \alpha_6 A_6(n) - \alpha_{11}A_7(n))A_7(q) +
   A_{11}(p)A_6(q) ({\rm i}\alpha_6A_5(n) \\
   &+ \alpha_{11}A_8(n)\Big)\Big]|0\rangle_x |aux\rangle_y |1\rangle_{ph}\\
&+\alpha_{12}|\psi^{(m)}\rangle_l|\psi^{(n)}\rangle_l|\psi^{(p)}\rangle_l|\psi^{(q)}\rangle_l|1\rangle_x |aux\rangle_y |1\rangle_{ph}
\end{split}
\end{equation*}
\begin{bfseries}
\flushleft{[State after step 5]}
\end{bfseries}
\begin{equation*}
\begin{split}
&|\psi^{(1)}\rangle=\Big(\alpha_1 A_1(m) A_1(r) + \alpha_3 A_1(r) A_4(m)
+ (\alpha_1A_2(m) - \alpha_3A_3(m)) A_4(r)\Big)\\
&~~~~~~~~|\psi^{(n)}\rangle_l|\psi^{(p)}\rangle_l|\psi^{(q)}\rangle_l|0\rangle_x|0\rangle_y |0\rangle_{ph}\\
&+A_1(m)A_1(r)\Big[a5[p]A_5(q)\Big(\alpha_2 A_5(n) - {\rm i} \alpha_7A_8(n)\Big)\\
&-A_9(p)\Big(\alpha_2A_6(n) + {\rm i}\alpha_7A_7(n)\Big)A_8(q)\Big]\\
&+A_4(r)|\psi^{(p)}\rangle_l\Big[A_2(m)\Big(\alpha_2 A_5(n)A_5(q)-{\rm i} \alpha_7 A_5(q) A_8(n)\\
&-\alpha_2A_6(n)A_8(q)-
{\rm i} \alpha_7A_7(n)A_8(q)\Big)\\
&+A_3(m)\Big(-\alpha_4 A_5(n) A_5(q)+{\rm i}\alpha_9 A_5(q)A_8(n)\\
& + \alpha_4 A_6(n)A_8(q)+{\rm i}\alpha_9 A_7(n)A_8(q)\Big)\Big]\\
&+ A_1(r) \Big[-\alpha_5 A_{12}(p) |\psi^{(m)}\rangle_l|\psi^{(n)}\rangle_lA_8(q)\\
&+A_4(m)\Big(|\psi^{(p)}\rangle_lA_5(q) (\alpha_4A_5(n) -{\rm i}\alpha_9A_8(n))\\
    & - A_9(p)(\alpha_4 A_6(n) +{\rm i}\alpha_9A_7(n))A_8(q)\Big)\Big]|1\rangle_x|0\rangle_y |0\rangle_{ph}\\
&+\Big(\alpha_1 A_1(m)A_2(r) - \alpha_1 A_2(m)A_3(r) + \alpha_3A_3(m)A_3(r)+ \alpha_3A_2(r)A_4(m)\Big)\\
&~~~~~~~~~~~|\psi^{(n)}\rangle_l|\psi^{(p)}\rangle_l|\psi^{(q)}\rangle_l|0\rangle_x|1\rangle_y |0\rangle_{ph}\\
&+\Big\{\alpha_4 A_3(m) A_3(r) |\psi^{(p)}\rangle_l A_5(n)A_5(q) + \alpha_4 A_2(r)A_4(m)|\psi^{(p)}\rangle_l |\psi^{(n)}\rangle_l|\psi^{(q)}\rangle_l\\
&- {\rm i}\alpha_9 A_3(m)A_3(r)|\psi^{(p)}\rangle_l A_5(q)A_8(n) -
 {\rm i}\alpha_9  A_2(r)A_4(m)|\psi^{(p)}\rangle_l|\psi^{(q)}\rangle_lA_8(n)\\
  &-\alpha_5 A_{12}(p)A_2(r)|\psi^{(m)}\rangle_l|\psi^{(n)}\rangle_l A_8(q) - \alpha_4 A_9(p)A_2(r) A_4(m)A_6(n) A_8(q)\\
  & - \alpha_4 A_3(m)A_3(r) A_5(p)A_6(n) A_8(q)- {\rm i}\alpha_9  A_9(p)A_2(r)\times\\
  & \times A_4(m)A_7(n) A_8(q) -
 {\rm i}\alpha_9  A_3(m)A_3(r)|\psi^{(p)}\rangle_l\times\\
 &\times A_7(n)A_8(q) +
  A_2(m)A_3(r) |\psi^{(p)}\rangle_l(-\alpha_2 A_5(n)A_5(q)\\
  &+ {\rm i}\alpha_7 A_5(q)A_8(n) + \alpha_2  A_6(n)A_8(q) +
 {\rm i}\alpha_7A_7(n)A_8(q))\\
 &+A_1(m)A_2(r)\Big[|\psi^{(p)}\rangle_l A_5(q)\Big(\alpha_2A_5(n)\\
 &-{\rm i}\alpha_7 A_8(n)\Big) - A_9(p)\Big(\alpha_2 A_6(n) + {\rm i}\alpha_7 A_7(n)\Big)A_8(q)\Big]\Big\}|1\rangle_x|1\rangle_y |0\rangle_{ph}\\
&+|\psi^{(q)}\rangle_l|\psi^{(r)}\rangle_l\Big\{\alpha_5A_{11}(p)|\psi^{(m)}\rangle_l|\psi^{(n)}\rangle_l -
   A_{10}(p)\Big[A_{1}(m)\Big(\alpha_2A_{6}(n)+ {\rm i}\alpha_7A_{7}(n)\Big)\\
   \end{split}
\end{equation*}
\begin{equation*}
\begin{split}
   & + A_{4}(m)(\alpha_4 A_{6}(n) + {\rm i}\alpha_9A_{7}(n))\Big]\Big\}|0\rangle_x|aux\rangle_y |0\rangle_{ph}\\
&-{\rm i}|\psi^{(r)}\rangle_l\Big[\alpha_8 A_1(m)  A_{10}(p)|\psi^{(n)}\rangle_lA_5(q) \\
&+ \alpha_{10} A_{10}(p)  A_4(m)|\psi^{(n)}\rangle_l A_5(q) \\
& +|\psi^{(m)}\rangle_l\Big( A_{11}(p) A_5(q) ({\rm i}\alpha_6 A_5(n) + \alpha_{11} A_8(n)) +
     |\psi^{(p)}\rangle_l(-{\rm i}\alpha_6 A_6(n)\\
      &+\alpha_{11}  A_7(n))  A_8(q)\Big)\Big]|1\rangle_x|aux\rangle_y |0\rangle_{ph}\\
&-{\rm i}\Big\{A_1(m)A_1(r)\Big[A_9(p) \Big(\alpha_2A_6(n) + {\rm i} \alpha_7 A_7(n)\Big) A_7(q) \\
&+ |\psi^{(p)}\rangle_lA_6(q)(\alpha_2A_5(n) -{\rm i}\alpha_7 A_8(n))\Big]\\
      & +A_1(r)\Big[\alpha_5 A_{12}(p)|\psi^{(m)}\rangle_l|\psi^{(n)}\rangle_lA_7(q)\\
   & +A_4(m)\Big(A_9(p)(\alpha_4A_6(n)+ {\rm i}\alpha_9 A_7(n))A_7(q)+
         |\psi^{(p)}\rangle_l\\
         &+A_6(q)(\alpha_4A_5(n)-{\rm i}\alpha_9 A_8(n))\Big)\Big] +
  A_4(r)|\psi^{(p)}\rangle_l\Big[A_2(m)(\alpha_2A_5(n)A_6(q)\\
  &+ \alpha_2 A_6(n)A_7(q)+
        {\rm i} \alpha_7 (A_7(n)A_7(q) - A_6(q)A_8(n)))\\
        & -A_3(m)\Big(\alpha_4A_5(n)A_5(q)+ \alpha_4A_6(n)A_7(q) \\
        &+{\rm i}\alpha_9 (A_7(n)A_7(q) - A_6(q)A_8(n))\Big)\Big]\Big\}|0\rangle_x|0\rangle_y |1\rangle_{ph}\\
&+|\psi^{(q)}\rangle_l \Big\{\alpha_8A_1(m)A_1(r)A_9(p) |\psi^{(n)}\rangle_l\\
&+ \Big(\alpha_8A_2(m) - \alpha_{10} A_3(m)\Big)A_4(r)|\psi^{(n)}\rangle_l|\psi^{(p)}\rangle_l\\
& +A_1(r)\Big[\alpha_{10} A_9(p)A_4(m)|\psi^{(n)}\rangle_l\\
&+A_{12}(p)|\psi^{(m)}\rangle_l\Big(-{\rm i}\alpha_6A_5(n)- \alpha_{11}A_8(n)\Big)\Big]\Big\}|1\rangle_x|0\rangle_y |1\rangle_{ph}\\
&-{\rm i}\Big\{\alpha_4 A_ 3 (m) A_ 3 (r)|\psi^{(p)}\rangle_l A_ 5 (n) A_6 (q) +
\alpha_4 A_ 2 (r) A_ 4 (m) |\psi^{(p)}\rangle_l A_ 5(n) A_6(q) \\
&+\alpha_5 A_{12} (p) A_ 2 (r)|\psi^{(m)}\rangle_l|\psi^{(n)}\rangle_l A_ 7(q) \\
&+ \alpha_4 A_ 9 (p) A_ 2
(r) A_ 4 (m) A_ 6 (n) A_ 7(q) + \alpha_4 A_ 3 (m) A_ 3 (r) |\psi^{(p)}\rangle_l\\
&A_ 6(n) A_ 7 (q) + {\rm i}\alpha_9 A_ 9 (p) A_ 2 (r) A_ 4 (m) A_ 7(n) A_ 7(q) \\
&+  {\rm i}\alpha_9 A_ 3 (m) A_ 3 (r) |\psi^{(p)}\rangle_l A_ 7 (n) A_7 (q)\\
   & -
   {\rm i}\alpha_9 A_ 3 (m) A_ 3 (r)|\psi^{(p)}\rangle_l A_ 6 (q) A_ 8 (n)\\
   & - {\rm i}\alpha_9 A_ 2 (r) A_ 4 (m) |\psi^{(p)}\rangle_l A_6 (q) A_ 8 (n)\\
    \end{split}
\end{equation*}
\begin{equation*}
\begin{split}
   & +A_ 1 (m) A_ 2 (r)\Big[A_ 9 (p) (\alpha_2 A_ 6 (n) +
        {\rm i}\alpha_6 A_ 7 (n)) A_7 (q) \\
        &+
    |\psi^{(p)}\rangle_l A_ 6 (q)\Big(\alpha_2 A_ 5(n) - {\rm i}\alpha_7 A_ 8(n)\Big)\Big]-A_ 2 (m) A_ 3 (r)|\psi^{(p)}\rangle_l\times \\
    &\times \Big[\alpha_2 A_ 5(n) A_6(q) +
\alpha-2 A_ 6 (n) A_ 7 (q) \\
&+{\rm i}\alpha_7\Big (A_ 7(n) A_ 7 (q) - A_ 6 (q) A_ 8 (n)\Big)\Big]\Big\}|0\rangle_x|1\rangle_y |1\rangle_{ph}\\
&+|\psi^{(q)}\rangle_l\Big(\alpha_8 A_ 1 (m) A_ 9(p) A_ 2 (r)|\psi^{(n)}\rangle_l + \alpha_{10}
A_ 9 (p) A_ 2 (r) A_ 4 (m)|\psi^{(n)}\rangle_l \\
&- \alpha_8 A_ 2 (m) A_ 3 (r)|\psi^{(n)}\rangle_l |\psi^{(p)}\rangle_l \\
 &+ \alpha_{10} A_ 3 (m) A_ 3 (r)|\psi^{(n)}\rangle_l|\psi^{(p)}\rangle_l\\
& -{\rm i}\alpha_6 A_ {12} (p) A_ 2 (r)|\psi^{(m)}\rangle_l A_ 5 (n) - \alpha_{11} A_
{12}(p) A_ 2 (r) |\psi^{(m)}\rangle_lA_ 8 (n)\Big)|1\rangle_x|1\rangle_y |1\rangle_{ph}\\
&-|\psi^{(r)}\rangle_l\Big\{\alpha_8 A_ 1 (m) A_ {10} (p) |\psi^{(n)}\rangle_l A_ 6 (q) \\
&+
\alpha_{10} A_ {10} (p) A_ 4 (m)|\psi^{(n)}\rangle_l A_ 6 (q) +
   |\psi^{(m)}\rangle_l\Big[|\psi^{(p)}\rangle_l ( {\rm i} \alpha_6 A_ 6 (n) - \alpha_{11} A_ 7 (n))\\
   & A_7(q) + A_ {11} (p) A_ 6 (q)\Big( {\rm i}\alpha_6 A_ 5(n) + \alpha_{11} A_ 8
(n)\Big)\Big]\Big\}|0\rangle_x|aux\rangle_y |1\rangle_{ph}\\
&+\alpha_{12}|\psi^{(m)}\rangle_l|\psi^{(n)}\rangle_l|\psi^{(p)}\rangle_l|\psi^{(q)}\rangle_l|\psi^{(r)}\rangle_l|1\rangle_x|aux\rangle_y |1\rangle_{ph}.
\end{split}
\end{equation*}

\end{document}